\documentclass[11pt]{article} 
\usepackage{mystyle-new}
\usepackage{epsfig,amsmath} 
\usepackage{graphics}
\usepackage{hyperref}
\usepackage{cite,./mcite}
\usepackage{listings}
\usepackage{color}
\usepackage{booktabs}
\usepackage{tabularx}
\usepackage{tikz}

\usepackage{lineno}

 \newcommand{\arrowcom}[1]{\textcolor{red}{\textbf{$\Longrightarrow$ #1}} \\}

\def\??#1{\mbox{}\\\arrowcom{#1}}
 
\pdfoutput=1

  \newenvironment{defl}[1]%
  {\begin{list}{}{\settowidth{\labelwidth}{#1}%
  \setlength{\leftmargin}{\labelwidth}%
  \addtolength{\leftmargin}{\labelsep}%
  \setlength{\itemsep}{0pt plus 1pt}
  \setlength{\parsep}{0pt plus 1pt}
  \setlength{\topsep}{0pt plus 1pt}
  \setlength{\partopsep}{0pt plus 1pt}
  \setlength{\parskip}{2mm plus 1mm minus 1mm}
  }}%
  {\end{list}}

\definecolor{listinggray}{gray}{0.9}
\definecolor{lbcolor}{rgb}{0.9,0.9,0.9}
\def\lsim{\mathrel{\rlap{\lower4pt\hbox{\hskip1pt$\sim$}}
    \raise1pt\hbox{$<$}}}                
\def\gsim{\mathrel{\rlap{\lower4pt\hbox{\hskip1pt$\sim$}}
    \raise1pt\hbox{$>$}}}                

\newcommand{\as}{\alpha_\mathrm{s}}

\def\kt{\ensuremath{k_t}}

\def\tmdlib{{TMDlib}}  
\def\tmdplotter{{TMDplotter}}  

\newcommand{\PBM}{PB}

\newenvironment{tolerant}[1]{\par\tolerance=#1\relax}{ \par }

\begin{document}

\begin{flushright}
DESY 21-026\\
IFJPAN-IV-2021-4\\
JLAB-THY-21-3337
\end{flushright}
\begin{center} {\sffamily\Large\bfseries \tmdlib 2  and \tmdplotter : \\
a platform for 3D hadron structure studies
 }
\\ \vspace{0.5cm}
{ 
 \Large 
N.A.~Abdulov$^{1}$, 
A.~Bacchetta$^{2}$, 
S.~Baranov$^{3}$, 
A.~Bermudez~Martinez$^{4}$, 
V.~Bertone$^{5}$,
C.~Bissolotti$^{2,6}$,
V.~Candelise$^{7}$, 
L.I.~Estevez~Banos$^{4}$,
M.~Bury$^{8}$, 
P.L.S.~Connor$^{4,}\footnote{Now at University of Hamburg}$,
L.~Favart$^{9}$,
F.~Guzman$^{10}$, 
F.~Hautmann$^{11,12}$,
M.~Hentschinski$^{13}$,
H.~Jung$^{4}$,
L.~Keersmaekers$^{11}$, 
A.~Kotikov$^{14}$,
A.~Kusina$^{15}$,
K.~Kutak$^{15}$,
A.~Lelek$^{11}$,
J.~Lidrych$^{4}$,
A.~Lipatov$^{1}$, 
G.~Lykasov$^{14}$,
M.~Malyshev$^{1}$,
M.~Mendizabal$^{4}$, 
S.~Prestel$^{16}$,
S.~Sadeghi~Barzani$^{17,11}$, 
S.~Sapeta$^{15}$,
M.~Schmitz$^{4}$,
A.~Signori$^{2,18}$,
G.~Sorrentino$^7$,
S.~Taheri~Monfared$^{4}$,
A.~van~Hameren$^{15}$,
A.M.~van~Kampen$^{11}$, 
M.~Vanden~Bemden$^{9}$,
A.~Vladimirov$^{8}$,
Q.~Wang$^{4,19}$,
H.~Yang$^{4,19}$
}\\  \vspace*{0.1cm}
{
    {$^{1}$SINP, Moscow State University, Russia }\\
    {$^{2}$Dipartimento di Fisica, Universit\`a di Pavia and INFN, Italy}\\
     {$^{3}$Lebedev Physics Institute, Russia }\\
      {$^4$DESY, Hamburg, Germany}\\
      {$^{5}$IRFU, CEA, Universit\'e Paris-Saclay, Gif-sur-Yvette, France }\\      
      {$^{6}$EIC Center, Jefferson Lab, Newport News, USA}\\
     {$^{7}$INFN Sezione di Trieste (a), Universit\`a di Trieste (b) , Trieste, Italy}\\
      {$^{8}$Institut f\"ur Theoretische Physik, Universit\"at Regensburg, Germany}\\
      {$^{9}$Inter-University Institute For High Energies, Universite Libre de Bruxelles, Belgium}\\
     {$^{10}$InSTEC, Universidad de La Habana, Cuba}\\
    {$^{11}$Elementary Particle Physics, University of Antwerp, Belgium}\\
    {$^{12}$RAL and University of Oxford, UK} \\
    {$^{13}$Departamento de Actuaria, Fisica y Matem\'aticas, Universidad de las Am\'ericas, Puebla, Mexico} \\
    {$^{14}$JINR, Dubna, Russia }\\
    {$^{15}$Institute of Nuclear Physics, Polish Academy of Sciences, Cracow, Poland} \\
    {$^{16}$Department of Astronomy and Theoretical Physics, Lund University, Sweden}\\
    {$^{17}$Department of Physics, Shahid Beheshti University, Iran}\\
   {$^{18}$Theory Center, Jefferson Lab, Newport News, USA}\\
    {$^{19}$School of Physics, Peking University, China} 
\vspace*{0.6cm}
}
\end{center}

\begin{abstract}
\noindent 
A common library, \tmdlib 2, for Transverse-Momentum-Dependent distributions (TMDs) and unintegrated parton distributions (uPDFs) is described, which allows for easy access of commonly used TMDs and uPDFs,  providing a three-dimensional (3D) picture of the partonic structure of hadrons.
The tool \tmdplotter\ allows for web-based plotting of distributions implemented in \tmdlib 2, together with collinear pdfs as available in LHAPDF.
\end{abstract} 

\newpage

\section*{PROGRAM SUMMARY}
\label{sec:summary}
{\em Computer for which the program is designed and others on which it is operable:}   any with standard C++, tested on Linux and Mac OS systems \\ \\
{\em Programming Language used:}  C++ \\ \\
{\em High-speed storage required:}  No \\ \\
{\em Separate documentation available: } No \\ \\
{\em Keywords: } QCD, TMD factorization, high-energy factorization, TMD PDFs, TMD FFs, unintegrated PDFs, small-$x$ physics.\\ \\
{\em Other programs used:}  LHAPDF (version 6) for access to collinear parton distributions, {\sc Root} (any version $>$ 5.30) for plotting the results \\ \\
{\em Download of the program:} \verb+http://tmdlib.hepforge.org+ \\ \\
{\em Unusual features of the program:}   None \\ \\
{\em Contacts:}   H. Jung (hannes.jung@desy.de), A. Bermudez Martinez (armando.bermudez.martinez@desy.de)
\\ \\
{\em Citation policy:} please cite the current version of the manual and the paper(s) related to the parameterization(s). \\
\newpage

\section{Introduction}
\label{sec:Introduction}

The calculation of processes at high energy hadron colliders is based in general on the calculation of a partonic process (matrix element) convoluted with the likelihood to find a parton of specific flavor and momentum fraction at a given scale within the hadrons. If the parton density depends only on the longitudinal momentum fraction $x$ of the hadron's momentum carried by a parton, and the resolution scale $\mu$, the processes are described by collinear factorization with the appropriate evolution of the parton densities (PDFs) given by the Dokshitzer-Gribov-Lipatov-Altarelli-Parisi (DGLAP) evolution equations~\cite{Gribov:1972ri,Altarelli:1977zs,Dokshitzer:1977sg}.  Such descriptions are successful for sufficiently inclusive processes, like inclusive deep-inelastic lepton-hadron scattering (DIS).

In several less inclusive processes, also the transverse momentum of the interacting partons plays an important role, leading to an extension of the collinear factorization theorem to include transverse degrees of freedom.  Different factorization theorems for the inclusion of transverse momenta to the parton densities have been developed in the past, leading to so-called Transverse Momentum Dependent (TMD) parton densities and unintegrated parton densities (uPDFs)~\cite{Angeles-Martinez:2015sea}.  These densities provide a 3D imaging of hadron structure, extending the 1D picture given by PDFs.
For semi-inclusive processes, like semi-inclusive DIS (SIDIS), Drell-Yan (DY) production and $e^+e^-$ scattering, TMD factorization has been formulated~\cite{Collins:1981uk,Collins:1981uw,Collins:1982wa,Collins:1981tt,Collins:1984kg,Collins:2011zzd,Meng:1995yn,Nadolsky:1999kb,Nadolsky:2000ky,Ji:2004wu,Ji:2004xq,GarciaEchevarria:2011rb,Chiu:2011qc}. The high-energy (small-$x$ limit) factorization was formulated for heavy flavor and heavy boson production in Refs.~\cite{Catani:1990xk,Catani:1990eg,Levin:1991ry,Collins:1991ty,Catani:1993ww,Catani:1994sq,Hautmann:2002tu} using unintegrated gluon distributions~\cite{Avsar:2012hj,Avsar:2011tz,Jadach:2009gm,Dominguez:2011saa,Dominguez:2011br,Dominguez:2011gc,Hautmann:2009zzb,Hautmann:2012pf,Hautmann:2007gw}. In Refs.~\cite{Hautmann:2017fcj,Hautmann:2017xtx} the Parton Branching (\PBM ) method was formulated as a way to obtain TMD distributions for all flavours over a wide range of $x$, transverse momentum $\kt$, and scale $\mu$ 
essentially by solving next-to-leading-order (NLO) DGLAP equations through Sudakov form factors, separating resolvable and 
non-resolvable branchings via the notion of soft-gluon resolution scale \cite{webber:1986mc, Ellis:1991qj}, and keeping track of the transverse momenta at each branching.

Since the number of available TMD densities increases very rapidly, and different groups provide different sets, it was necessary to develop a common platform to access the different TMD sets in a common form. 
In 2014 the first version of \tmdlib\ (version 1) and \tmdplotter\ was released \cite{Hautmann:2014kza,Connor:2016bmt}, which made several TMD sets available to the community. This library has set a common standard for accessing TMD sets, similar to what was available for collinear parton densities in PDFlib~\cite{PlothowBesch:1992qj,PlothowBesch:1995ci} and LHAPDF~\cite{Buckley:2014ana}.
\tmdlib\ is a {\tt C++} library which provides a framework and an  interface to a collection of different uPDF and TMD parameterizations.

In this report, we describe a new version of the TMDlib library, \tmdlib 2, as well 
as the associated online plotting tool \tmdplotter. \tmdlib 2 covers all the features present 
already in the previous  version and contains significant new developments, such as 
the treatment of TMD uncertainties and a more efficient method to include new TMD sets. 
The report is structured as follows. In Sec.~2, we give the main elements of the library 
framework. In Sec.~3 we emphasize the new features of \tmdlib 2 compared to the 
previous version. In Sec.~4 we provide the essential documentation. We summarize in Sec.~5. 

\section{The TMDlib framework\label{sec2}}

The TMDlib library and its new version \tmdlib 2 consider 
momentum weighted TMD parton distributions 
$x {\cal A}_j (x ,\kt, \mu)$ of flavor $j$ as functions of the parton's  light cone longitudinal momentum fractions $x$ of the hadron's momentum, the parton's  transverse momentum \kt , and the evolution scale $\mu$ \cite{Angeles-Martinez:2015sea}.
Besides, the library also contains integrated TMDs obtained from the integration over $\kt$, as follows  
\begin{equation}
\label{intA}
x{\cal A}_{int}(x,\mu) = \int_{k_{t,min}}^{k_{t,max}}d \kt^2  \;  x{\cal A}(x,\kt , \mu) \, ,
\end{equation}

\begin{figure}[htb]
\begin{center}
\includegraphics[width=0.49\textwidth]{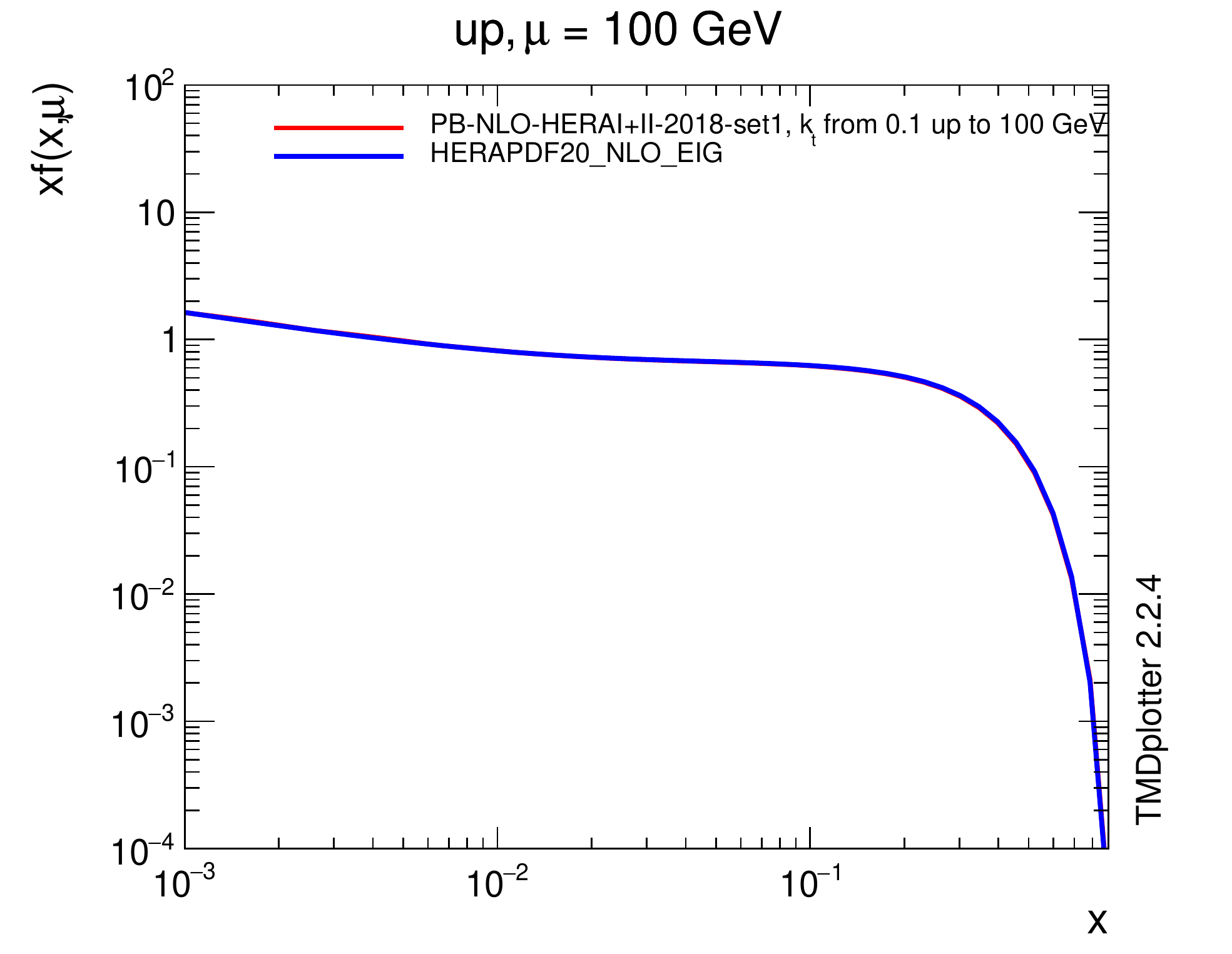}
\includegraphics[width=0.49\textwidth]{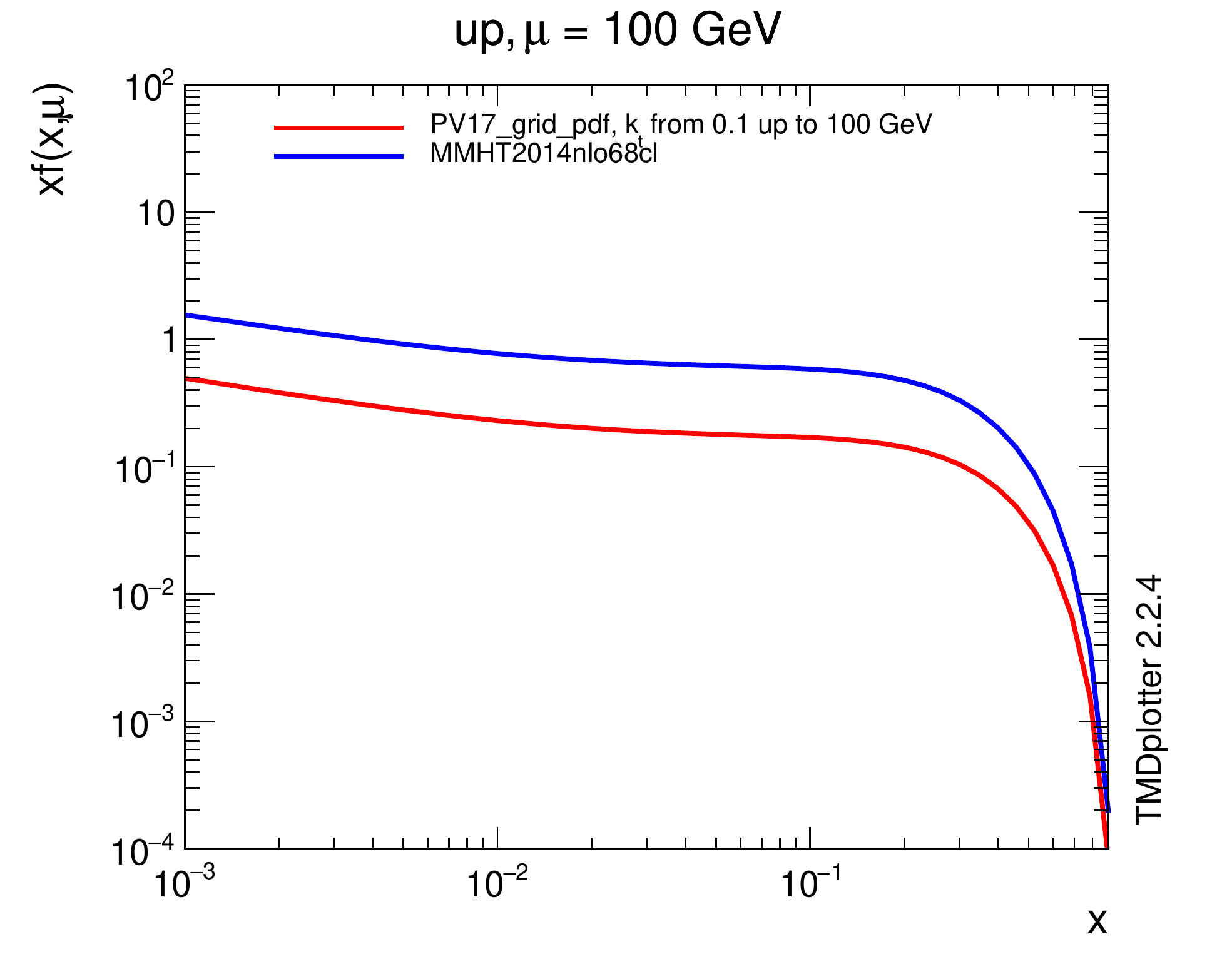}
\caption{Comparison of up-type parton distributions, $xf(x,\mu)=x{\cal A}_{int}(x,\mu)$ as a function of $x$ at $\mu=100$ GeV 
(left): comparison of 
the integrated distribution PB-NLO-HERAI+II-2018-set1~\protect\cite{Martinez:2018jxt} with HERAPDF2.0~\protect\cite{Abramowicz:2015mha}. 
(right): comparison of integrated distribution PV17~\protect\cite{Bacchetta:2017gcc} with MMHT2014~\protect\cite{Harland-Lang:2014zoa}.
}
\label{fig:TMDplottercollinear} 
\end{center}
\end{figure}

In Fig.~\ref{fig:TMDplottercollinear} (left), we show an example of integrated TMD obtained with \tmdplotter\ 
for the PB-NLO-HERAI+II-2018-set1 \cite{Martinez:2018jxt}, in which the integral between $k_{t,min} =0.01$ and $k_{t,max} =100$  GeV is 
compared with the collinear PDF set HERAPDF2.0 \cite{Abramowicz:2015mha}. By construction both sets are identical. 
However, in general, Eq.(\ref{intA}) does not converge to the collinear pdf, which is shown in Fig.~\ref{fig:TMDplottercollinear} (right) comparing the integral between $k_{t,min} =0.01$ and $k_{t,max} =100$  GeV  of PV17~\cite{Bacchetta:2017gcc} with the corresponding collinear distribution of MMHT2014~\protect\cite{Harland-Lang:2014zoa}. 
Several aspects of the relationship between integrated TMDs and collinear PDFs have been investigated in the literature. The matching coefficient between the integrated gluon TMD and the collinear gluon PDF in the MSbar scheme was first computed in the small-$x$ limit in Ref.~\cite{Catani:1993ww}, with small-$x$ resummation of logarithmic accuracy $(\alpha_s \ln x )^m$ to all orders $m$ in $\alpha_s$. Perturbative calculations of the matching coefficients at finite order have recently been carried out through N$^3$LO in Refs.~\cite{Ebert:2020yqt,Luo:2019szz}. Other aspects of the relationship between integrated TMDs and collinear PDFs are studied e.g. in \cite{Catani:1990eg,Watt:2003mx,Hautmann:2007uw,Hautmann:2006xc,Avsar:2012hj,Avsar:2011tz,Collins:2007ph}. We refer the  reader to the overview \cite{Angeles-Martinez:2015sea}, and references therein, for further discussions of this topic. 

In \tmdlib 2 the  densities are defined more generally as momentum weighted distributions $x{\cal A}(x,\bar{x}, \kt, \mu)$, where $x, \bar{x}$ are the (positive and negative) light-cone longitudinal momentum fractions~\cite{Watt:2003vf,Watt:2003mx,Collins:2005uv,Collins:2007ph}.  In some of the applications $\bar{x}$ is set explicitly to zero, while in other cases $\bar{x} = 0$ means that it is implicitly integrated over.

\subsection{Grids and Interpolation}
Since the analytic calculation of TMDs as a function of the longitudinal momentum fraction $x$ (we neglect $\bar{x}$ in the following), the transverse momentum \kt\ and the scale $\mu$ is very time consuming and in some cases even not available, the TMDs are saved as grids, and \tmdlib\ provides appropriate tools for interpolation between the grid points (where the type of evolution is indicated):

\begin{center}
\begin{tabularx}{\textwidth}{p{0.2\textwidth}X}
\toprule
allFlavPDF & Multidimensional Linear Interpolation in $x$, \kt\ and $\mu$ is used for \PBM\ and CCFM-type TMDs.  \\
{Pavia} & Interpolation based on Lagrange polynomials of degree three, performed through APFEL++\cite{Bertone:2017gds,Bertone:2013vaa}.\\ 
{InterpolationKS} & Multidimensional cubic spline interpolation in $x$, $k_t$ and $\mu$, based on GSL implementation, is used for KS-type TMDs (see Tab.~\ref{tab:TMDuPDF_sets}). \\
\bottomrule
\end{tabularx}
\end{center}
The parameterizations of TMDs in \tmdlib\ are explicitly authorized for each distribution by the corresponding authors.
A list of presently available TMD sets is given in Tab.~\ref{tab:TMDuPDF_sets}.
No explicit QCD evolution code is included: the parameterizations are as given in the corresponding references.

The grids of each selected TMD set are read into memory once (the I/O time depends on the size of the grid). Each TMD set is initialized as a separate instance of the TMD class,  which is created for each different TMD set, for example for uncertainty sets, or if several different TMD sets are needed for the calculation. 
The memory consumption of TMDlib is determined by the size of the TMDgrids. Optionally, TMDgrids can be loaded separately, avoiding large memory consumption.

It is the philosophy of \tmdlib\, that the definition of TMD grids is left free, but a few examples are given:
the grids for the \PBM, CCFM and KS TMD sets are stored in form of text tables, the grids of the Pavia type TMDs are stored and read via the \verb+YAML+ frame. 
The method of interpolation and the corresponding accuracy of the interpolation is left under control of the authors of the relevant TMD sets.
\begin{figure}[htb]
\begin{center}
\includegraphics[width=0.49\textwidth]{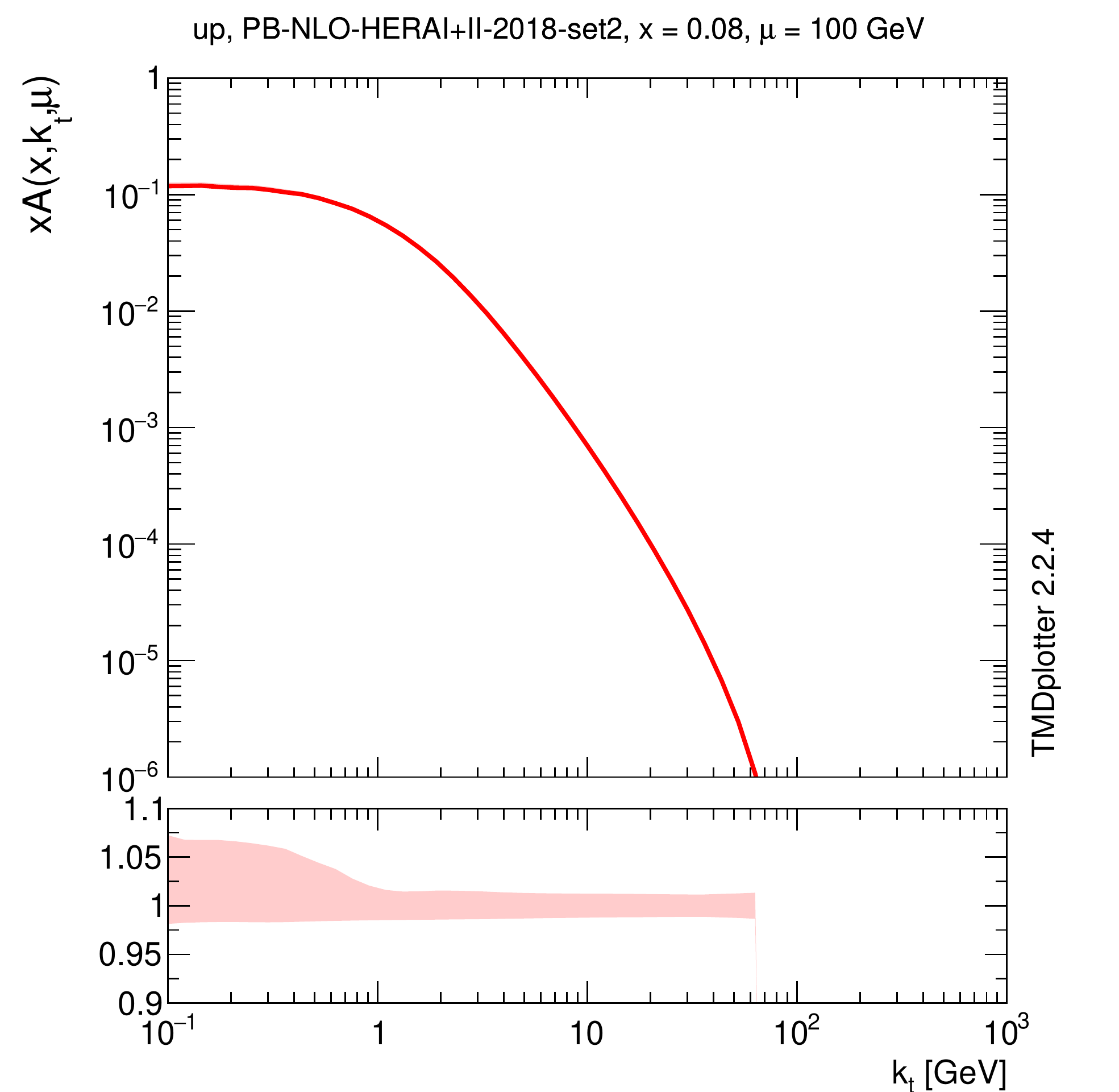}
\includegraphics[width=0.49\textwidth]{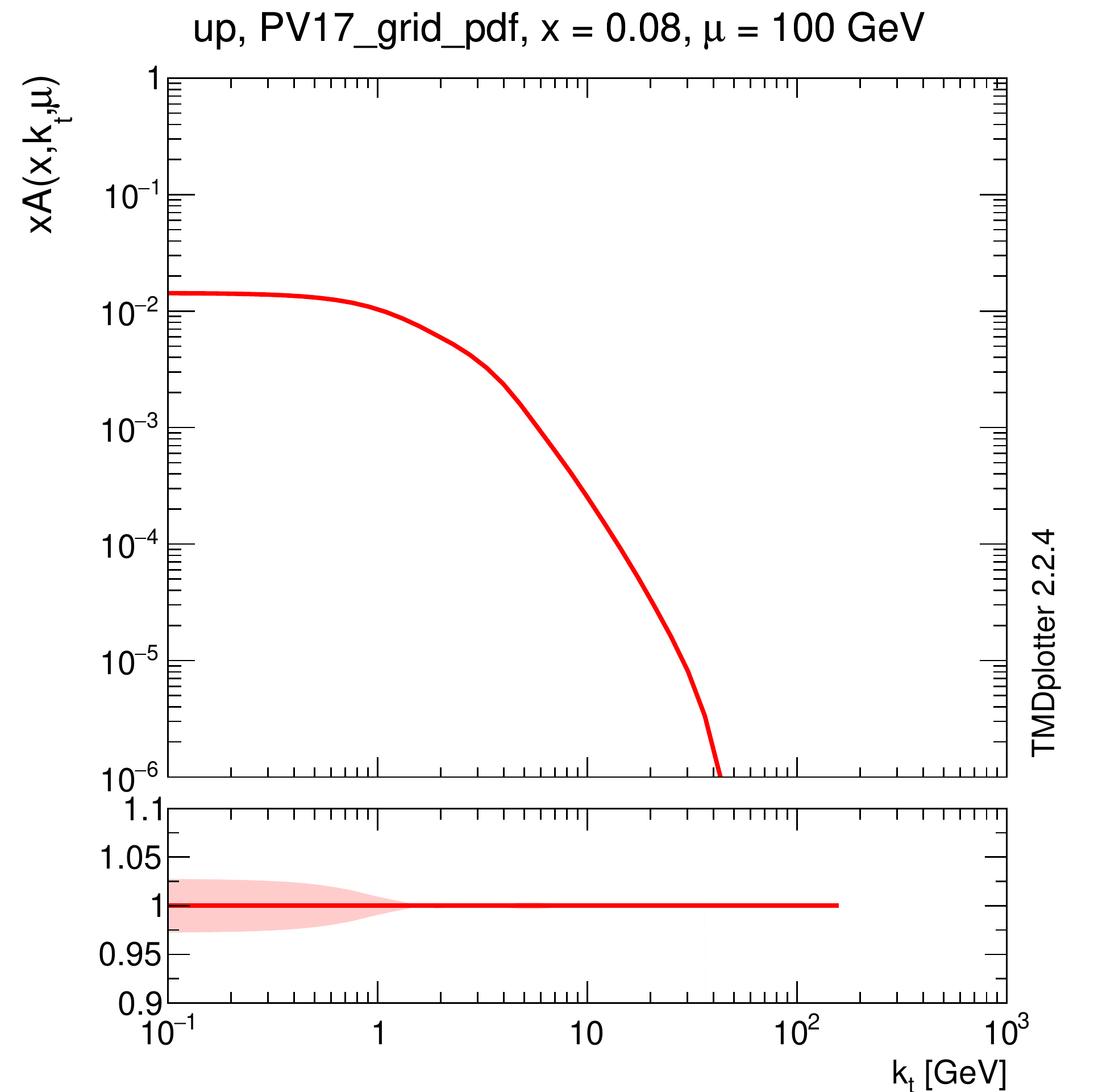}
\caption{Transverse momentum distribution $x{\cal A}(x,\kt,\mu)$ at $x=0.08$ and $\mu=100$ GeV obtained with
PB-NLO-HERAI+II-2018-set2~\protect\cite{Martinez:2018jxt} (left) and 
 PV17~\protect\cite{Bacchetta:2017gcc} (right). }
\label{fig:TMDuncertainty} 
\end{center}
\end{figure}

\subsection{Uncertainty TMD sets}
The estimation of theoretical uncertainties is an important ingredient for phenomenological applications, and uncertainties from PDFs and TMDs play a central role.
The uncertainties of TMDs are estimated usually from the uncertainties of the input parameters or parameterization. There are two different methods commonly used: the Hessian method~\cite{Pumplin:2002vw} which is applied if the parameter variations are orthogonal  or the Monte Carlo method  providing Monte Carlo replicas~\cite{Giele:1998gw,Giele:2001mr}.  
The specific prescriptions on how to calculate the uncertainties for a given TMD set should be found in the original publication describing the TMDs.

An example of TMDs with uncertainty band is shown in Fig.~\ref{fig:TMDuncertainty} for the \PBM\ set as well as for the PV17 set.
The parameters of intrinsic \kt - distribution are part of the fit of PV17, while they are not fitted for the  \PBM\  sets (see discussion in Ref.~\cite{Martinez:2018jxt}).

\subsection{TMDplotter}
\begin{figure}[htb]
\begin{center}
\includegraphics[width=0.49\textwidth]{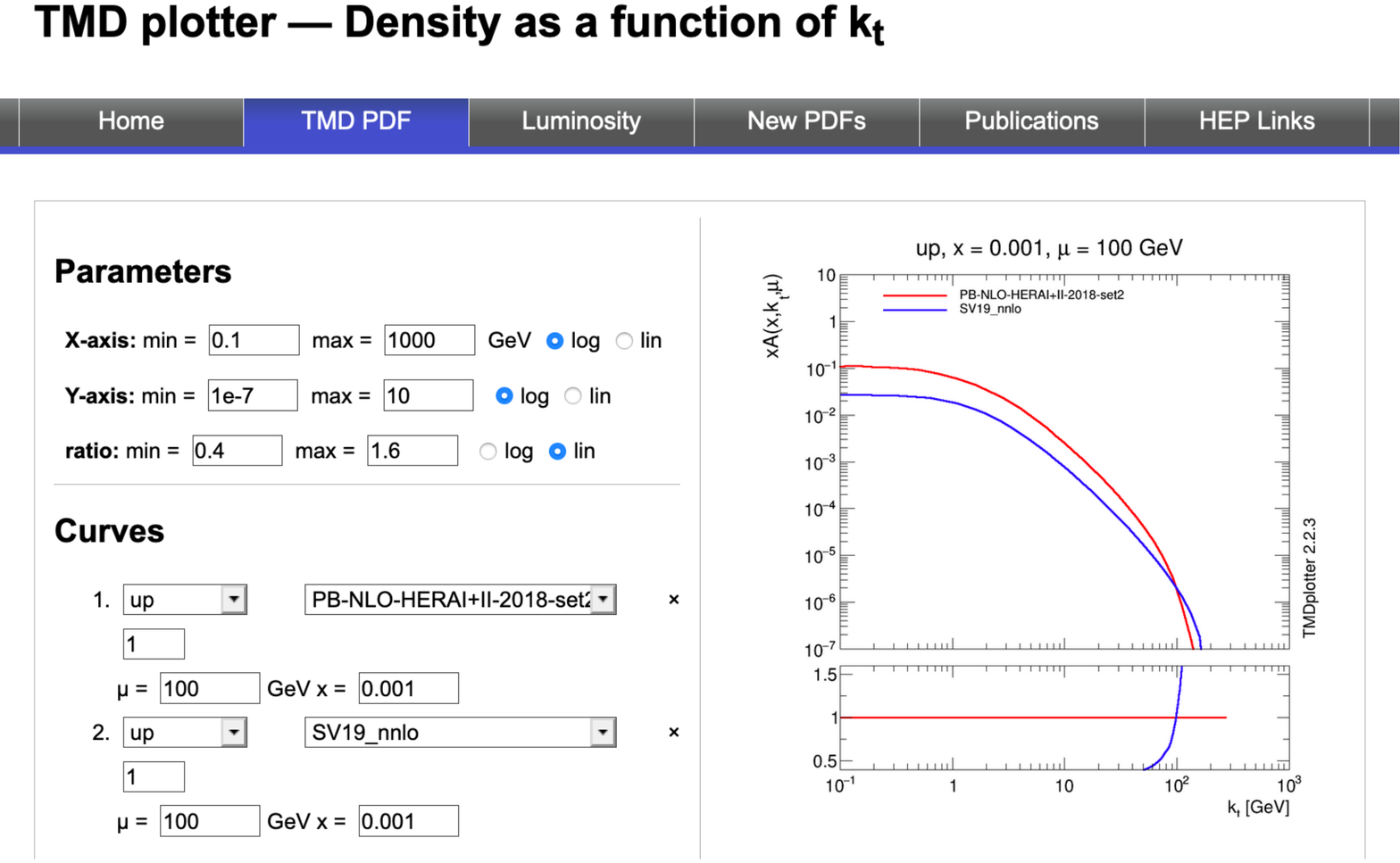}
\includegraphics[width=0.46\textwidth]{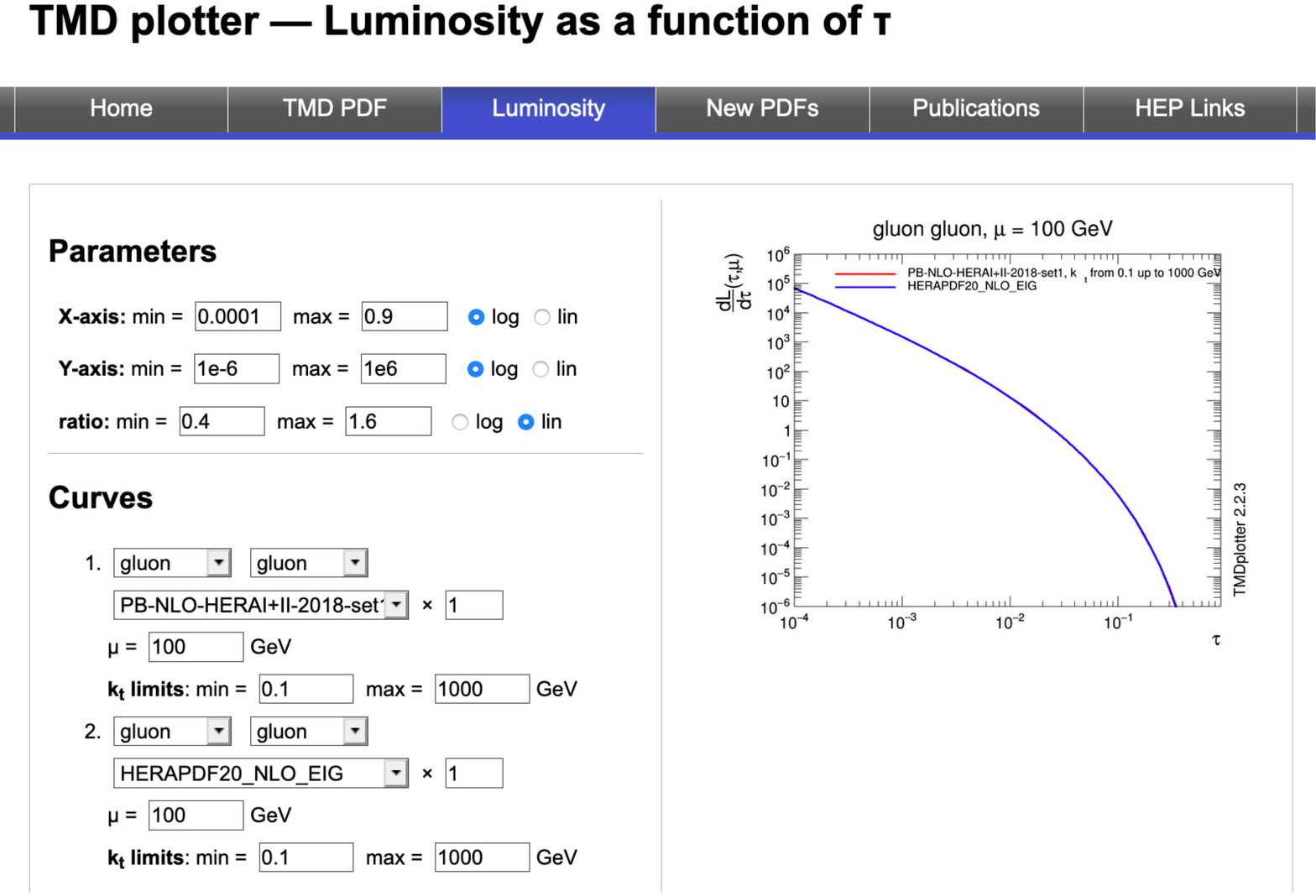}
\caption{\protect\tmdplotter\  examples: (left) comparison of the transverse momentum distributions of different TMD sets, (right) gluon-gluon luminosity calculation using integrated TMD sets (the curves from PB-NLO-HERAI+II-2018-set1 and  {\tt HERAPDF2.0} overlap).
}
\label{fig:TMDplotter} 
\end{center}
\end{figure}

\begin{tolerant}{2000}
\tmdlib\ provides also a web-based application for plotting TMD distributions -- \tmdplotter , plotting tools for collinear pdfs are available under e.g. \cite{ApfelWeb} or \cite{DurhamPlotter}.
In Fig.~\ref{fig:TMDplotter} (left) a comparison of the transverse momentum distributions of different TMD sets is shown, and in Fig.~\ref{fig:TMDplotter} (right) the gluon-gluon luminosity calculation for the integrated TMD sets PB-NLO-HERAI+II-2018-set1~\protect\cite{Martinez:2018jxt} at $\mu=100$~GeV compared with the one obtained from {\tt HERAPDF2.0} is shown (the curves obtained from PB-NLO-HERAI+II-2018-set1 and  {\tt HERAPDF2.0} overlap).
\end{tolerant}

\tmdplotter\ is available at \url{http://tmdplotter.desy.de/}. 

\section{New features}
Having described in Sec.~\ref{sec2} the general framework of the \tmdlib\  library, we here stress the main new features of \tmdlib 2 compared with the previous version \cite{Hautmann:2014kza} of the library.  The most important development concerns the inclusion of many new TMD sets. This is achieved through a  new and more efficient method to add input files. The method is flexible enough that it will allow new sets, which may become available in the future, to also be included in a straightforward manner.   Another extremely important development of \tmdlib 2, which plays an essential role in paving the way to systematic TMD phenomenology at colliders and fixed target machines, is that the uncertainties associated with TMD sets are now accessible through the library. This was not the case in the first version \cite{Hautmann:2014kza}. It is the first time that TMD uncertainties become available in a library tool. While uncertainties on collinear PDFs are nowadays available through several different web-based resources, \tmdlib 2 is at present the unique tool which contains the full existing information on uncertainties on TMD sets, and makes it readily accessible. As such, we expect it to be an essential tool for phenomenological studies of TMDs and comparisons with experimental data.        

To sum up, the main new features of \tmdlib 2 compared 
to the earlier version of the library are as follows.

\begin{itemize}
\item \tmdlib 2 makes use of C++ classes, and the different sets corresponding to uncertainty sets or sets corresponding to different parameterizations are read once and  initialized as different instances, allowing to load many sets into memory;
\item information about TMD sets is read via {\tt YAML} from the TMD info files, containing all metadata;
\item including new TMD sets is simplified with the new structure of the input sets;
\item the TMD sets are no longer part of the TMDlib distribution, but can be downloaded via {\tt TMDlib-getdata}, distributed with \tmdlib 2.
\end{itemize}
\section{{\tmdlib} documentation}
\label{sec:TMDlibdocumentation}

{\tmdlib} is written in {\tt C++} , with an interface for access from {\tt FORTRAN} code.
The source code of {\tmdlib} is available from {\url{http://tmdlib.hepforge.org/}~} and can be installed using the \textit{standard} {\tt autotools} sequence {\tt configure}, {\tt make},  {\tt make install}, with options to specify the installation path and the location of the {\tt LHAPDF} PDF library~\cite{Buckley:2014ana},  and the {\tt ROOT} data analysis framework library~\cite{Brun:1997pa} (which is used optionally for plotting). If {\tt ROOT} is not found via {\tt root-config}, the plotting option is disabled. After installation, {\tt TMDlib-config} gives access to necessary environment variables.

\subsection{Description of the program components }


\subsubsection*{Initialization in C++}
{\small
\begin{center}
 \begin{tabularx}{\textwidth}{p{0.35\textwidth}X}
  \toprule
 {\tt TMDinit(name)} & To initialize the dataset specified by its name {\tt name}.
                      A complete list of datasets available in the current version of
                      {\tmdlib} with the corresponding name is provided in 
                      Tab.~\ref{tab:TMDuPDF_sets}. \\
   \midrule
   {\tt TMDinit(name,irep)} &
                       To initialize a given {\tt irep} replica of the dataset  {\tt name}.\\
   \midrule 
   {\tt TMDinit(iset)} & To initialize the dataset specified by its identifier {\tt iset}. \\
  \bottomrule
 \end{tabularx}
\end{center}
\subsubsection*{Initialization in Fortran}
\begin{center}
 \begin{tabularx}{\textwidth}{p{0.35\textwidth}X}
  \toprule
   {\tt TMDinit(iset)} & To initialize the dataset specified by its identifier {\tt iset}. \\
   \midrule 
   {\tt TMDset(iset)} & To switch to the dataset {\tt iset}. \\
  \bottomrule
 \end{tabularx}
\end{center}
}
\subsubsection*{Access to TMDs in C++}
{\small
 \begin{tabularx}{\textwidth}{p{0.35\textwidth}X}
  \toprule
  {\small \tt TMDpdf(x, xbar, kt, mu)} &  

                        Vector double-type function returning an array of $13$ variables for QCD parton densities
                       with the values of $x{\cal A}(x,{\bar x},\kt,\mu)$:
                       at index $0,\dots,5$ is $\bar{t},\dots,\bar{d}$, at index $6$
                       is the gluon, and at index $7,\dots,12$ is $d,\dots,t$ densities. \\
   \midrule
    {\small \tt TMDpdf(x, xbar, kt, mu, xpq)}&  
                        Void-type function filling an array of $13$ variables, {\tt xpq},
                       with the values of $x{\cal A}(x,{\bar x},\kt,\mu)$ :
                       at index $0,\dots,5$ is $\bar{t},\dots,\bar{d}$, at index $6$
                       is the gluon, and at index $7,\dots,12$ is $d,\dots,t$ densities.\\  
   \midrule
    {\small \tt TMDpdf(x, xbar, kt, mu, uval, dval, sea, charm, bottom, gluon, photon)} &                      
                       Void-type function to return $x{\cal A}(x,\bar{x},k_t,\mu)$ at $x, \bar{x}, k_t,\mu $ 
                       for valence u-quarks {\tt uval}, valence d-quarks { \tt dval}, light sea-quarks {\tt s}, 
                       charm-quarks {\tt c}, bottom-quarks {\tt b}, gluons {\tt glu} and  gauge boson   {\tt photon}. \\
   \midrule
    {\small \tt TMDpdf(x, xbar, kt, mu, up, ubar, down, dbar, strange, sbar, charm, cbar, bottom, bbar, gluon, photon)} &  
                        To return $x{\cal A}(x,\bar{x},k_t,\mu)$ at $x, \bar{x}, k_t,\mu $ for the partons \verb+up+, \verb+ubar+, \verb+down+, \verb+dbar+, \verb+strange+,
                       \verb+sbar+, \verb+charm+, \verb+cbar+, \verb+bottom+, \verb+bbar+, \verb+gluon+ and gauge boson  \verb+photon+ (if available). \\ 
    \midrule
    {\small \tt TMDpdf(x, xbar, kt, mu, up, ubar, down, dbar, strange, sbar, charm, cbar, bottom, bbar, gluon, photon, Z0, W+,W-,higgs)} & 
                        To return $x{\cal A}(x,\bar{x},k_t,\mu)$ at $x, \bar{x}, k_t,\mu $ for the partons \verb+ up+, \verb+ubar+, \verb+down+, \verb+dbar+, \verb+strange+,
                       \verb+sbar+, \verb+charm+, \verb+cbar+, \verb+bottom+, \verb+bbar+, \verb+gluon+, the  gauge bosons \verb+photon+,
                       \verb+Z0+, \verb+W++, \verb+W-+ and \verb+higgs+ (if available). \\ 
  \bottomrule
 \end{tabularx}
\subsection*{Access to TMDs in Fortran}
 \begin{tabularx}{\textwidth}{p{0.35\textwidth}X}
  \toprule
    {\small \tt TMDpdf(kf,x, xbar, kt, mu, up, ubar, down, dbar, strange, sbar, charm, cbar, bottom, bbar, gluon)} &  
                        To return $x{\cal A}(x,\bar{x},k_t,\mu)$ at $x, \bar{x}, k_t,\mu $ for the partons \verb+up+, \verb+ubar+, \verb+down+, \verb+dbar+, \verb+strange+,
                       \verb+sbar+, \verb+charm+, \verb+cbar+, \verb+bottom+, \verb+bbar+, \verb+gluon+ for the hadron flavor \verb+kf+.
                       (\verb+kf+ is no longer used, only kept for backward compatibility with TMDlib1)\\ 
    \midrule
    {\small \tt TMDpdfEW(x, xbar, kt, mu, up, ubar, down, dbar, strange, sbar, charm, cbar, bottom, bbar, gluon, photon, Z0, W+,W-, higgs)} & 
                        To return $x{\cal A}(x,\bar{x},k_t,\mu)$ at $x, \bar{x}, k_t,\mu $ for the partons \verb+ up+, \verb+ubar+, \verb+down+, \verb+dbar+, \verb+strange+,
                       \verb+sbar+, \verb+charm+, \verb+cbar+, \verb+bottom+, \verb+bbar+, \verb+gluon+, the  gauge bosons \verb+photon+,
                       \verb+Z0+, \verb+W++, \verb+W-+ and \verb+higgs+ (if available). \\ 
  \bottomrule
 \end{tabularx}
  }
\subsubsection*{Callable program components}
The program components listed in this section are accessible with the same name in C++ as well as in Fortran.
{\small
\begin{center}
 \begin{tabularx}{\textwidth}{p{0.35\textwidth}X}
  \toprule
  {\tt TMDinfo(dataset)}   & Accesses information from the \verb+info+ file.\\
  \midrule
  {\tt TMDgetDesc()}   & Returns data set description from \verb+info+ file.\\
  \midrule
  {\tt TMDgetIndex()}   & Returns index number as a string of data set from \verb+info+ file.\\
  \midrule
  {\tt TMDgetNumMembers()}   & Returns number of members of data sets from \verb+info+ file. \\
  \midrule
  {\tt TMDgetScheme()}   & Returns evolution scheme of dataset from \verb+info+ file.\\
  \midrule
  {\tt TMDgetNf()}   & Returns the number of flavours, $N_f$, used for the computation of $\Lambda_{QCD}$.\\
  \midrule
  {\tt TMDgetOrderAlphaS()} & Returns the perturbative order of $\as$ used in the evolution of 
                      the dataset.\\
  \midrule
  {\tt TMDgetOrderPDF()} 
                      & Returns the perturbative order of the evolution of the dataset.\\ 
  \midrule                  
  {\tt TMDgetXmin()}  & Returns the minimum value of the momentum fraction $x$ 
                      for which the dataset initialized by {\tt TMDinit(name)} was determined.\\
  \midrule
  {\tt TMDgetXmax()}  & Returns the maximum value of the momentum fraction $x$
                      for which the dataset initialized by {\tt TMDinit(name)} was determined.\\
  \midrule
  {\tt TMDgetQmin()}  ({\tt TMDgetQ2min()})  & Returns the minimum value of the energy scale $\mu$ (in GeV), ($\mu^2$ (in GeV$^2$))
                     for dataset.\\  
  \midrule
   {\tt TMDgetQmax()}  ({\tt TMDgetQ2max()}) & Returns the maximum value of the energy scale $\mu$ (in GeV) , ($\mu^2$ (in GeV$^2$))
                      for dataset.\\
  \midrule
  {\tt TMDgetExtrapolation\_Q2()} 
                      & Returns the method of extrapolation in scale  outside the grid definition as specified in  \verb+info+ file.\\ 
  \midrule
  {\tt TMDgetExtrapolation\_kt()} 
                      & Returns the method of extrapolation in $\kt$ outside the grid definition as specified in  \verb+info+ file.\\ 
  \midrule
  {\tt TMDgetExtrapolation\_x()} 
                      & Returns the method of extrapolation in $x$ outside the grid definition as specified in  \verb+info+ file.\\ 
  \midrule
  {\tt TMDnumberPDF(name)}  & Returns the identifier as a value of the associated \verb+name+ of the dataset.\\                 
 \midrule
  {\tt TMDstringPDF(index)}  & Returns the name associated with  \verb+index+ of the dataset.\\                 
  \bottomrule
 \end{tabularx}
\end{center}
}

\subsection{{\tmdlib} calling sequence}

In the following simple examples are given to demonstrate how information from the TMD parton densities can be obtained in C++ and Fortran.

\begin{small}
\begin{itemize}
\item in C++ 
\begin{verbatim}
      string name ="PB-NLO-HERAI+II-2018-set2";
      double x=0.01, xbar=0, kt=10., mu=100.;
      TMD TMDtest;
      int irep=0;
      TMDtest.TMDinit(name,irep);
      cout << "TMDSet Description: " << TMDtest.TMDgetDesc() << endl;
      cout << "number       = " << TMDtest.TMDnumberPDF(name) << endl;
      TMDtest.TMDpdf(x,xbar,kt,mu, up, ubar, down, dbar, strange, sbar, 
               charm, cbar, bottom, bbar, gluon, photon);
\end{verbatim}
\item in Fortran (using multiple replicas of the TMD)
\begin{verbatim}
      x = 0.01
      xbar = 0
      kt = 10.
      mu = 100.
      iset = 102200
      call TMDinit(iset)
      write(6,*) ' iset = ', iset
      call TMDinit(iset)
      nmem=TMDgetNumMembers()
      write(6,*) ' Nr of members ', nmem,' in Iset = ', iset
      do i=0,nmem
        isetTMDlib = iset+i 
        write(6,*) ' isetTMDlib = ', isetTMDlib
        call TMDinit(isetTMDlib)
        call TMDset(isetTMDlib)
        call TMDpdf(kf,x,xbar,kt,mu,up,ubar,dn,dbar,strange,sbar,
    &    charm,cbar,bottom,bbar,glu)
        call TMDpdfew(kf,x,xbar,kt,mu,up,ubar,dn,dbar,strange,sbar,
    &    charm,cbar,bottom,bbar,glu,photon,z0,wplus,wminus,higgs)
      end do
     
\end{verbatim}
\end{itemize}
\end{small}

\subsection{Installation of TMD grids }
The TMD grid files are no longer automatically distributed with the code package, but have to be installed separately.
A list of available TMD parameterizations is given in Tab.~\ref{tab:TMDuPDF_sets}.

\begin{small}
\begin{verbatim}
# get help
bin/TMDlib-getdata --help

# install all data sets
bin/TMDlib-getdata all 

# install a single data (for example: SV19_nnlo)
bin/TMDlib-getdata SV19_nnlo

\end{verbatim}
\end{small}

\subsection{Structure of TMD grids}
In \tmdlib 2 the TMDgrids are stored in directories with the name of a given TMD set which is located in {\tt installation\_prefix/share/tmdlib/TMDsetName}. Every such directory contains info file and grid file(s), for example for a TMD set called \verb+test+:
\begin{small}
\begin{verbatim}
~/local/share/tmdlib> ls test
test.info     test_0000.dat 
\end{verbatim}
\end{small}
\begin{tolerant}{2000}
\noindent
The \verb+info+ file contains general information on the TMDset (inspired by a similar strategy in LHAPDF), as described below, and the file(s) \verb+test_0000.dat+ contains the TMDgrid. If further replicas are available (for example for uncertainties), the files are numbered as  \verb+test_0000.dat, test_0001.dat,...+, with the number of files given by \verb+NumMembers+ as described below.
\end{tolerant}

\noindent
The \verb+info+ file must contain all the information to initialize and use the TMDgrid:
\begin{small}
\begin{verbatim}
SetDesc: "Description of the dataset "
SetIndex: XXXXX
Authors: XXXX
Reference: XXXX
Particle: 2212
NumMembers: 34
NumFlavors: 6
TMDScheme: PB TMD
Flavors: [-5, -4, -3, -2, -1, 1, 2, 3, 4, 5, 21]
Extrapolation_x: fixed
Extrapolation_Q2: fixed
Extrapolation_kt: fixed
AlphaS_MZ: 0.118
AlphaS_OrderQCD: 1
OrderQCD: 1
XMin: 9.9e-07
XMax: 1.
KtMin: 0.01
KtMax: 13300.
QMin: 1.3784
QMax: 13300
MZ: 91.1876
MUp: 0.
MDown: 0.
MStrange: 0.
MCharm: 1.47
MBottom: 4.5
MTop: 173
\end{verbatim}
\end{small}
The meaning of most entries is obvious from their name, with \verb+TMDScheme+
different structures for the TMD grids can be selected:
\begin{defl}{1234567890123456789}
\item[{\tt PB TMD}] used for the PB TMD series
\item[{\tt PB TMD-EW}] used for the PB TMD series including electroweak particles
\item[{\tt Pavia TMDs}] used for the PaviaTMD (or similar TMD) series
\end{defl}

\begin{table}[htb]
\footnotesize
\centering
\begin{tabular}{r | l | c | c }
\toprule
iset		& uPDF/TMD set      & Subsets  & Ref. \\ \hline 
101000	&\verb+ccfm-JS-2001+				&1	&\cite{Jung:2004gs} \\
101010	&\verb+ccfm-setA0+				&	4	&\cite{Jung:2004gs} \\
101020	&\verb+ccfm-setB0+				&	4	&\cite{Jung:2004gs} \\
101001	&\verb+ccfm-JH-set1+				&1	&\cite{Hansson:2003xz} \\
101002	&\verb+ccfm-JH-set2+				&1	&\cite{Hansson:2003xz} \\
101003	&\verb+ccfm-JH-set3+				&1	&\cite{Hansson:2003xz} \\
101201	&\verb+ccfm-JH-2013-set1+		&		13	&\cite{Hautmann:2013tba} \\
101301	&\verb+ccfm-JH-2013-set2+		&		13	&\cite{Hautmann:2013tba} \\
101401     &\verb+MD-2018+ & 1 &  \cite{Abdulov:2018ccp}\\
101410     &\verb+KLSZ-2020+ & 1 & \cite{Kotikov:2019kci}\\ 
102100	&\verb#PB-NLO-HERAI+II-2018-set1#	&	35	&\cite{Martinez:2018jxt} \\	
102200	&\verb#PB-NLO-HERAI+II-2018-set2#&		37	&\cite{Martinez:2018jxt} \\
102139	&\verb#PB-NLO-HERAI+II-2018-set1-q0#	&	3	&\cite{Martinez:2018jxt} \\
102239	&\verb#PB-NLO-HERAI+II-2018-set2-q0#	&	3	&\cite{Martinez:2018jxt} \\		
103100  	&\verb#PB-NLO+QED-set1-HERAI+II#&	1	&\cite{Jung:2021mox}\\
103200  	&\verb#PB-NLO+QED-set2-HERAI+II #& 	1	&\cite{Jung:2021mox}\\				
10904300	&\verb+PB-NLO_ptoPb208-set1+			&1	&\cite{Blanco:2019qbm} \\
10904400	&\verb+PB-NLO_ptoPb208-set2+			&1	&\cite{Blanco:2019qbm} \\
10901300	&\verb+PB-EPPS16nlo_CT14nlo_Pb208-set1+	&1	&\cite{Blanco:2019qbm} \\
10901400	&\verb+PB-EPPS16nlo_CT14nlo_Pb208-set2+	&1	&\cite{Blanco:2019qbm} \\
10902300	&\verb+PB-nCTEQ15FullNuc_208_82-set1+		&33	&\cite{Blanco:2019qbm} \\
10902400	&\verb+PB-nCTEQ15FullNuc_208_82-set2+		&33	&\cite{Blanco:2019qbm} \\
200001	&\verb+GBWlight+				&	1	&\cite{GolecBiernat:1998js} \\
200002	&\verb+GBWcharm+				&	1	&\cite{GolecBiernat:1998js} \\
210001	&\verb+Blueml+					&1	&\cite{Blumlein:1995mi} \\
400001	&\verb+KS-2013-linear+			&	1	&\cite{Kutak:2012rf} \\
400002	&\verb+KS-2013-non-linear+			&1	&\cite{Kutak:2012rf} \\
400003	&\verb+KS-hardscale-linear+		&1	&\cite{Kutak:2014wga} \\
400004	&\verb+KS-hardscale-non-linear+		&	1	&\cite{Kutak:2014wga} \\
400101     &\verb+KS-WeizWill-2017+      		&   	1	& \cite{Kotko:2017oxg}\\
500001	&\verb+EKMP+				&		1	&\cite{Echevarria:2015uaa} \\
410001	&\verb+BHKS+			&		1	&\cite{Bury:2017jxo} \\
300001	&\verb+SBRS-2013-TMDPDFs+	&			1	&\cite{Signori:2013mda} \\
300002	&\verb+SBRS-2013-TMDPDFs-par+		&	1	&\cite{Signori:2013mda} \\
601000	&\verb+PV17_grid_pdf+	&		    	201	&\cite{Bacchetta:2017gcc} \\
602000	&\verb+PV17_grid_ff_Pim+	&		201	&\cite{Bacchetta:2017gcc} \\
603000	&\verb+PV17_grid_ff_Pip	+	&		201	&\cite{Bacchetta:2017gcc} \\
604000	&\verb+PV17_grid_FUUT_Pim+	&	100	&\cite{Bacchetta:2017gcc} \\
605000	&\verb+PV17_grid_FUUT_Pip+	&		100	&\cite{Bacchetta:2017gcc} \\
606000	&\verb+PV19_grid_pdf	+	&		216	&\cite{Bacchetta:2019sam} \\
607000 	&\verb+PV20_grid_FUTTsin_P_Pim+	&	101   & \cite{Bacchetta:2020gko}\\
608000 	&\verb+PV20_grid_FUTTsin_P_Pip+	&	101   & \cite{Bacchetta:2020gko}\\
701000 	&\verb+SV19_nnlo		+	&		23 	&\cite{Scimemi:2019cmh} \\
702000 	&\verb+SV19_nnlo_all=0 		+	&	21	&\cite{Scimemi:2019cmh} \\
703000 	&\verb+SV19_n3lo		+	&		23	&\cite{Scimemi:2019cmh} \\
704000 	&\verb+SV19_n3lo_all=0		+	&	21 	&\cite{Scimemi:2019cmh} \\
705000	&\verb+SV19_ff_pi_n3lo		+	&	23	&\cite{Scimemi:2019cmh} \\
706000	&\verb+SV19_ff_pi_n3lo_all=0	+	&	21	&\cite{Scimemi:2019cmh} \\
707000	&\verb+SV19_ff_K_n3lo		+	&	23	&\cite{Scimemi:2019cmh} \\
708000	&\verb+SV19_ff_K_n3lo_all=0	+	&	21	&\cite{Scimemi:2019cmh} \\
709000	&\verb+SV19_pion			+	&	7	&\cite{Vladimirov:2019bfa} \\
710000	&\verb+SV19_pion_all=0		+	&	7	&\cite{Vladimirov:2019bfa} \\
711000	&\verb+BPV20_Sivers		+	&	25	&\cite{Bury:2020vhj} \\
\bottomrule
\end{tabular}
\caption{Available uPDF/TMD parton sets in \tmdlib .}
\label{tab:TMDuPDF_sets}
\label{updfs}
\end{table}

\section{Summary}
\label{sec:conclusions}

The authors of this manual set up a collaboration to develop and maintain {\tmdlib} and {\tmdplotter}, respectively a {\tt C++} library for handling different parameterizations of uPDFs/ TMDs and a corresponding online plotting tool.
The aim is to update these tools with more uPDF/TMD parton sets and new features, as they 
become available and are developed.
\tmdlib 2 improves on the efficiency of previous versions, allows for simpler C++ interfaces and simplifies the inclusion of new uPDF/TMD sets.

\section*{Acknowledgments}
\begin{tolerant}{2000}
N. Abdulov was also supported by the RFBR grant 19-32-90096.
S. Baranov, A. Lipatov and M. Malyshev are grateful the DESY Directorate for the support in the framework of Cooperation Agreement between MSU and DESY on phenomenology of the LHC processes and TMD parton densities.
V.~Bertone is supported by the European Union's Horizon 2020 research and innovation programme under grant agreement No 824093.
C.~Bissolotti is supported by the  European Research Council (ERC) under the European Union's Horizon 2020 research and innovation program (grant agreement No. 647981, 3DSPIN) and by the U.S. Department  of  Energy  contract  DE-AC05-06OR23177 under  which  Jefferson  Science  Associates  operates  the Thomas Jefferson National Accelerator Facility.
F. Hautmann thanks DESY, Hamburg and CERN, Theory Group for hospitality and support.
M. Hentschinski gratefully acknowledges support by Consejo Nacional de Ciencia y Tecnolog\'ia grant number A1 S-43940 (CONACYT-SEP Ciencias B\'asicas). 
A. Lelek acknowledges funding by Research Foundation-Flanders (FWO) (application number: 1272421N).
M.~Malyshev and N.~Abdulov were supported by the grant of the Foundation for the Advancement of Theoretical Physics and Mathematics (grants 20-1-3-11-1 and 18-1-5-33-1, respectively.
A.~Signori acknowledges support from the European Commission through the Marie Sk\l{}odowska-Curie Action SQuHadron (grant agreement ID: 795475). 
S.~Taheri~Monfared thanks the Humboldt Foundation for the Georg Forster research fellowship  and gratefully acknowledges support from IPM.    
A.~van~Hameren acknowledges support from the Polish National Science Centre grant no. 2019/35/ST2/03531. K.~Kutak acknowledges the support by Polish National Science Centre grant no.  DEC-2017/27/B/ST2/01985
Q.~Wang and H.~Yang acknowledge the support by the Ministry of Science and Technology under grant No. 2018YFA040390 and by the National Natural Science Foundation of China under grant No. 11661141008. 

\end{tolerant}

\providecommand{\etal}{et al.\xspace}
\providecommand{\href}[2]{#2}
\providecommand{\coll}{Coll.}
\catcode`\@=11
\def\@bibitem#1{%
\ifmc@bstsupport
  \mc@iftail{#1}%
    {;\newline\ignorespaces}%
    {\ifmc@first\else.\fi\orig@bibitem{#1}}
  \mc@firstfalse
\else
  \mc@iftail{#1}%
    {\ignorespaces}%
    {\orig@bibitem{#1}}%
\fi}%
\catcode`\@=12

\end{document}